\documentclass{emulateapj}

\shorttitle{The Red Halo Phenomenon}
\shortauthors{Zackrisson et al.}

\begin{document}

\title{The Red Halo Phenomenon}

\author{E. Zackrisson\altaffilmark{1,2}, N. Bergvall\altaffilmark{1}, G. \"Ostlin\altaffilmark{3}, G. Micheva\altaffilmark{3}, \& M. Leksell\altaffilmark{1}}
\altaffiltext{1}{Department of Astronomy and Space Physics, Box 515, 751 20 Uppsala,
Sweden (ez@astro.uu.se, nils.bergvall@astro.uu.se, maria.leksell@home.se)}
\altaffiltext{2}{Tuorla Observatory, V\"ais\"antie 20, Piikki\"o, FIN-21500, Finland}
\altaffiltext{3}{Stockholm Observatory, AlbaNova University Center, 106 91 Stockholm, Sweden (ostlin@astro.su.se, genoveva@astro.su.se)}

\begin{abstract}
Optical and near-IR observations of the halos of disk galaxies and blue compact galaxies have revealed a very red spectral energy distribution which cannot easily be reconciled with a normal, metal-poor stellar population like that in the stellar halo of the Milky Way. Here, spectral evolutionary models are used to explore the consequences of these observations. We demonstrate that a stellar population of low to intermediate metallicity, but with an extremely bottom-heavy initial mass function, can explain the red halos around both types of objects. Other previously suggested explanations, like nebular emission or very metal-rich stars, are shown to fail in this respect. This indicates that, if the reported halo colours are correct, halo populations dominated by low-mass stars may be a phenomenon common to galaxies of very different Hubble types. Potential tests of this hypothesis are discussed, along with its implications for the baryonic dark matter content of galaxies. 
\end{abstract}





\section{Introduction}
To this day, the nature of galactic halos remains elusive. Apart from the known stellar constituents of the halo of the Milky Way, searches for Massive Astrophysical Compact Halo Objects (MACHOs) have reported the discovery of $\approx 0.1$--$1\ M_\odot$ objects of unknown origin \citep[e.g.][]{Alcock b, Calchi Novati}, which may potentially contribute substantially to the dark matter content of galaxies. 

In the quest to detect faint halo stars (possible MACHOs) around other galaxies during the mid-90s, attention gravitated towards the galaxy NGC 5907, as deep optical and near-IR images \citep[e.g.][]{Sackett et al.,Lequeux et al. a,Rudy et al.,James & Casali} indicated the detection of a very red halo population above the plane of its edge-on disk. While thick disks -- now believed to be common among disk galaxies -- can also extend far from the mid-plane and be quite red, their colours are still consistent with normal stellar populations \cite[][]{Dalcanton & Bernstein}, contrary to the colours reported for the halo of NGC 5907. Interest declined with the discovery of what appeared to be the remnants of a disrupted dwarf galaxy close to NGC 5907 \citep{Shang et al.} and suggestions that this feature, in combination with contamination by foreground stars, could have resulted in a spurious halo detection \citep{Zheng et al.}. Follow-up observations with the Hubble Space Telescope (HST) have, however, not been able to confirm this explanation for the strange properties of the integrated halo light \citep{Zepf et al.}. 

Recently, \citet*{Zibetti et al.} stacked the images of 1047 edge-on disk galaxies from the Sloan Digital Sky Survey (SDSS) and detected a halo population with a strong excess in the {\it i}-band and optical colours curiously similar to those previously derived for NGC 5907 -- again very difficult to reconcile with standard stellar populations. The halo detected around an edge-on disk galaxy at redshift $z=0.322$ in the Hubble Ultra Deep Field (HUDF) also shows puzzling colours \citep{Zibetti & Ferguson}.  

The red halo phenomenon is not necessarily confined to disk galaxies alone. Optical/near-IR broadband photometry of the faint halos of blue compact galaxies (BCGs) have similarly revealed a very red spectral energy distribution (\citealt{Bergvall & Östlin}; \citealt{Bergvall et al.}; \"Ostlin et al., in preparation), which cannot easily be reconciled with a metal-poor stellar population like that in the halo of the Milky Way. This red excess is not likely to be caused by dust reddening, given the upper limits on the presence of cold dust in BCG halos inferred from ISO data (Bergvall et al., in preparation), and the low extinction measured by the H$\alpha$/H$\beta$ Balmer decrement in the central starburst \citep{Bergvall & Östlin}. In the case of Haro 11, the BCG with the reddest halo observed so far, Bergvall \& \"Ostlin have furthermore used ISO observations to rule out near-IR emission from warm dust as an explanation for the red colours.

Are the red halos of disk galaxies and BCGs related, and if so -- what is the origin of the red excess? Previous investigations have suggested nebular emission \citep{Zibetti & Ferguson}, very metal-rich stars \citep{Rudy et al.,Bergvall & Östlin,Zibetti et al.},  or a stellar population dominated by low-mass stars \citep{Rudy et al.,Zepf et al.,Zibetti et al.} as possible explanations for the red colours. Here, we use spectral evolutionary models to test these different scenarios in detail. 

The red halo data used is briefly described in section 2. In section 3, we demonstrate that a stellar population with a low to intermediate metallicity and an extremely bottom-heavy initial mass function (IMF) simultaneously reproduces the colours of halos detected around disk galaxies and BCGs. In section 4, we show that although very metal-rich stars can explain the BCG halo colours, this solution fails for the red halos of disk galaxies. In section 5, nebular emission is shown to be much too blue to explain the stacked SDSS halo and the halos around BCGs, but does provide a possible explanation for the halo detected around a single disk galaxy in the HUDF. The implications of a halo population dominated by low-mass stars for the baryonic dark matter of galaxies  is discussed in section 6, along with a number of potential tests of this scenario.

\section{Red Halo Data}
For the halos detected around disk galaxies, we will use the $gri$ data for the halo detected in stacked SDSS data by \citet{Zibetti et al.}, and the $viz$ (F606W, F775W, F850LP) data for the halo of the highly inclined disk galaxy COMBO-17 31611 detected at a redshift of $z=0.322$ in the HUDF \citep{Zibetti & Ferguson}. For the BCGs, the $BVK$ halo data presented in \citet{Bergvall & Östlin} and \citet{Bergvall et al.}  will be used. 

It is important to realise that red halos may under certain circumstances result simply from instrumental effects \citep{Michard}. In the data discussed, careful examination of the point spread function does however rule this out as a significant contributor to the detected colours. Because of the suggestions that the light profiles of the NGC 5907 halo may in fact be affected by artefacts related to the point spread function \citep{Zheng et al.}, we refrain from any detailed analysis of the colours of this object. Nonetheless, we note that by combining the data from \citet{Rudy et al.} and \citet{Lequeux et al. b} for the east (more reliable) side of the galaxy, the existing profiles indicate $B-V\approx 1.0$, $V-K\approx 4.5$ at a distance of 4.3 kpc from the centre, which is fairly close to the colours of the reddest BCG halo in our sample (Haro 11, see Fig.~\ref{BCGfig}). 
 
\section{Low-Mass Stars}
Could it be that the red excess observed in the halos of disk galaxies and BCGs originates in a stellar population with a peculiar IMF? The most obvious way to produce a very red spectral energy distribution would be to increase the fraction of cool, low-mass stars by adopting a bottom-heavy IMF ($dN/dM\propto M^{-\alpha}$ with $\alpha>2.35$, where $\alpha=2.35$ represents the Salpeter slope). 

\begin{figure*}[t]
\resizebox{\hsize}{!}{\includegraphics{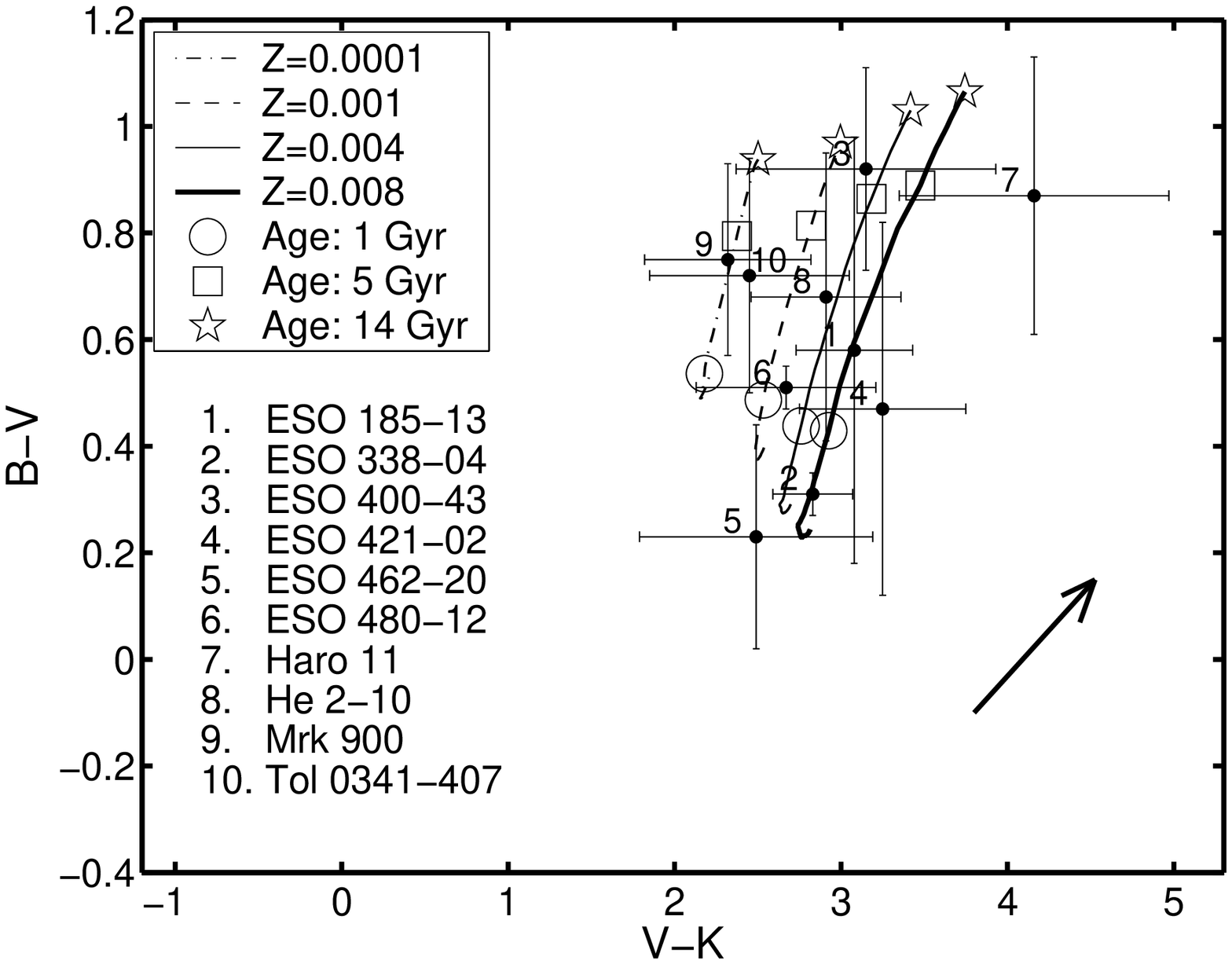}\includegraphics{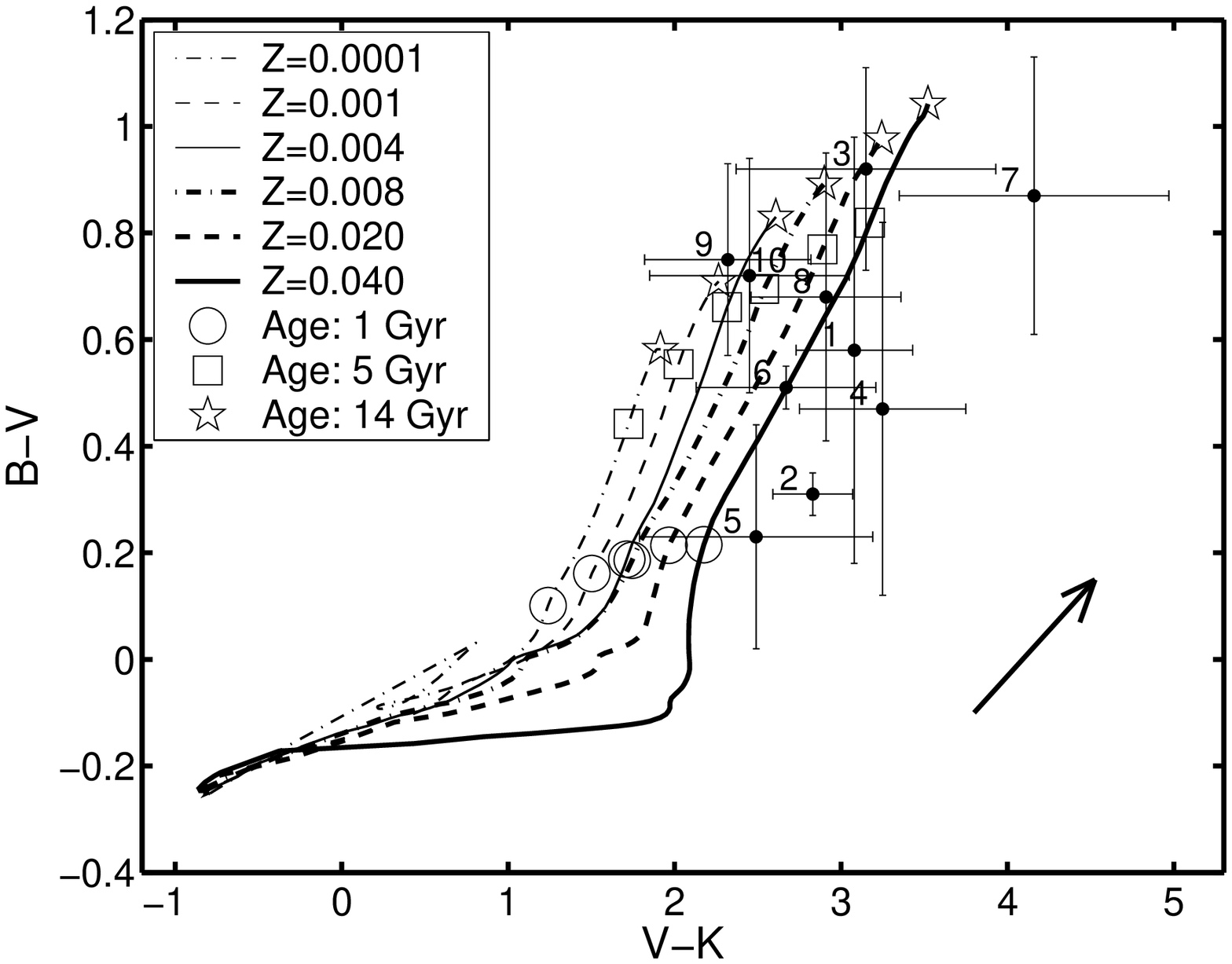}\includegraphics{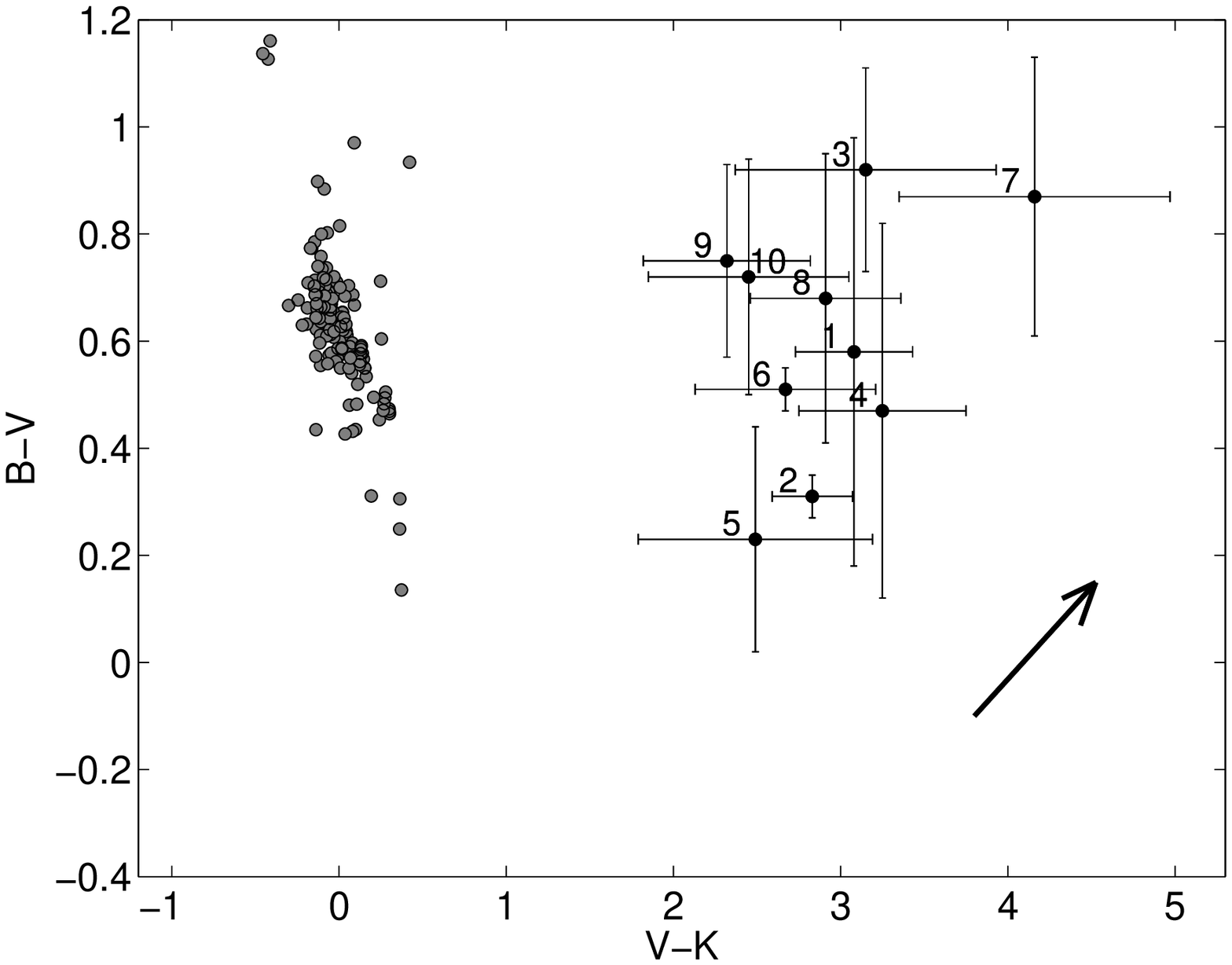}}
\caption{{\bf Left:} Observed colours of BCG halos (crosses indicating $1\sigma$ error bars), compared to the predictions of P\'EGASE.2 (lines) for stellar populations with a bottom heavy IMF ($dN/dM\propto M^{-\alpha}$ with $\alpha=4.50$) at different metallicities.
 With such an extreme IMF, stellar populations with low to intermediate metallicities ($Z=0.001$--0.008; thin and thick solid lines) provide a reasonable fit. An exponentially declining star formation rate (SFR$(t)\propto \exp{-t/\tau}$) with $\tau=1$ Gyr has been assumed. The arrow indicates the average Milky Way dust reddening vector \citep{Cardelli et al.} for $E(B-V)=0.25$. {\bf Middle: } Same, but for stellar populations with a Salpeter IMF. For many of the halos, only the highest metallicities ($Z\geq 0.020$; thick dashed and solid lines) provide a reasonable fit. {\bf Right: }Observed colours of BCG halos compared to the colours predicted for various lines of sight through photoionized nebulae (grey markers). These predictions indicate that nebular emission is much too blue in $V-K$ to explain the observed halo colours. \label{BCGfig}}
\end{figure*}

In Fig.~\ref{BCGfig} (left panel), we use the P\'EGASE.2 spectral evolutionary model \citep{Fioc & Rocca-Volmerange} to demonstrate that stellar populations with an extremely bottom-heavy IMF ($\alpha=4.50$) in fact can explain the BCG halo colours with low to intermediate stellar metallicities ($Z=0.001$--0.008). Although a star formation history (SFH) typical of an early-type galaxy (SFR$(t)\propto \exp(-t/\tau)$ with $\tau=1$ Gyr) is assumed here, variations in SFH (within reasonable limits) mainly affect the inferred age of the halo population (which in the $\tau=1$ Gyr scenario typically becomes 1--5 Gyr, although with substantial uncertainties). Acceptable fits to the halo colours can also be achieved for other SFHs and ages, as further discussed in Sect. 6. Interestingly, a stellar population with a similar age, metallicity and the same bottom-heavy IMF ($\alpha=4.50$) also succeeds in explaining the halo detected in stacked data edge-on disk galaxy data from the SDSS (Fig.~\ref{SDSSfig}, left panel).

As indicated by the dust reddening vector \citep[average Milky Way extinction curve, $R_V=3.1$, assumed;][]{Cardelli et al.} included in Fig.~\ref{BCGfig} \& \ref{SDSSfig}, the red halos around BCGs and edge-on disks can, on the other hand, not be explained by low-metallicity, Salpeter-IMF populations subject to dust reddening.

Since the uncertainties involved in spectral evolutionary synthesis are still fairly large, we have used the \citet[][ hereafter Z01]{Zackrisson et al.} spectral evolutionary model to check that these findings are not crucially dependent on any single model. We find that in all important aspects, Z01 confirms the bottom-heavy IMF conclusions obtained with P\'EGASE.2.

Here, the IMF is assumed to be a single-valued power-law throughout the $0.08\leq M/M_\odot\leq 120$ mass interval. Experimentation with model parameters does however indicate that for a stellar population with $\alpha=4.50$, a mass range of $0.10\leq M/M_\odot\leq 3$ ($0.15\leq M/M_\odot\leq 3$) is sufficient to reproduce the halo colours of BCGs with P\'EGASE.2 (Z01). To reproduce the SDSS halo colours, a mass span of $0.10\leq M/M_\odot\leq 3$ ($0.09\leq M/M_\odot\leq 2$) is required. The lower mass limit of the IMF is particularly important for the inferred mass-to-light ($M/L$) ratio of the halo population, as further discussed in section 6.

\section{Metal-rich Stars} 
In \citet{Bergvall & Östlin}, metal-rich stars were suggested as an explanation for the red excess of BCG halos. In Fig.~\ref{BCGfig}, we show that although a metal-rich population with a Salpeter IMF  does reasonably well when confronted with the extended BCG halo data set of \citet{Bergvall et al.}, the metallicities of many of the halos would have to be very high (solar or higher). Here, we use the P\'EGASE.2 model, although other models (Z01, \citealt{Bruzual & Charlot}) agree with this conclusion.

Such high stellar metallicities would be strange given the low metallicity ($\sim$10\% solar) of the gas in the central starburst of these objects. A possible explanation for this could be that BCGs are created through recent mergers between a metal-poor, gas-rich object (an intergalactic gas cloud or a low surface brightness galaxy) and a metal-rich, gas-poor elliptical galaxy. In this case, the detected halo would simply correspond to the elliptical host of the central starburst, which was ignited during the merger. This scenario would however require the BCG mass measurements based on gas dynamics \citep{Östlin et al.} to be substantial underestimates, either because of gas dissipation or insufficient time for the gas to relax enough to trace the underlying gravitational potential.

In Fig.~\ref{SDSSfig} (left panel), we use P\'EGASE.2 to show that stellar populations obeying a Salpeter IMF (dash-dotted lines) fail to explain the observed colours of the stacked SDSS halo, regardless of metallicity. The Z01 and \citet{Bruzual & Charlot} models agree with this conclusion. Since only Z01 allows colour predictions at non-zero redshifts, we use this model in Fig.~\ref{HUDF} to show that a metal-rich, Salpeter IMF population (thick dashed line) also fails to explain the colours of the HUDF halo.

\section{Nebular Emission}
In principle, the halo light could be affected by nebular emission originating in an extended envelope of gas ionized by hot stars in the central stellar component of these galaxies.  Since photoionization models predict the spectrum of a complete Str\"omgren sphere to be very blue, this explanation for the red excess in BCGs was dismissed by \citet{Bergvall & Östlin}. Current halo observations do not, however, probe the entire Str\"omgren sphere, but only lines of sight through the outer parts of the ionized cloud. Could it then be that these regions have a substantially different colour signature? 

\begin{table}[t]
\begin{flushleft}
\caption[]{The model parameter values used when modelling the nebular emission. All sequences assume a stellar component mass of $10^{10} \ M_\odot$. The generated grid consists of all possible combinations of the parameter values listed below.}
\begin{tabular}{ll} 
\tableline\tableline
$\log n(\mathrm{H})$ (cm$^{-3}$) & 0, 1, 2 \cr
Filling factor & 0.01, 0.1, 1.0 \cr
Gas metallicity, $Z $ & 0.001, 0.020$^1$ \cr
\tableline
\end{tabular}\\
$^1$ Only for the halos of disk galaxies\\
\label{paramtable}
\end{flushleft}
\end{table}

To model the spatially resolved spectral energy distribution of an ionized cloud is in principle a 3D problem, for which no suitable and publicly available photoionization currently exists. However, a (very) poor-man's 3D model can nonetheless be constructed, using results from the 1D photoionization model Cloudy version 90.05 \citep{Ferland et al.}. Here, a spectrum considered representative for the ionizing stellar population is used as input to Cloudy. For each such stellar population continuum, spherical nebulae with a wide range of densities, filling factors and metallicities are then generated, using all possible combinations of the parameter values listed in Table \ref{paramtable}. Every resulting model nebula is recomputed ten times, each time truncated at progressively smaller radii. From these results, the spectral energy distribution of each spherical shell is derived. The spectra of the nebular shells are finally weighted together according to their relative volumes along each of the possible lines of sight through the ionized cloud. This results in a large number of different line-of-sight model spectra, whose colours may be compared to those of the observed halos.

In the case of BCG halos, Z01 model spectra for metal-poor ($Z=0.001$) stellar populations, with constant star formation rates and ages of 10 Myr or 100 Myr, are used to represent the central starburst. In this case, the gas metallicity is assumed to be the same as that of the stars. The colours of the resulting line-of-sight models are compared to the BCG halo colours in Fig.~\ref{BCGfig} (right panel). Despite a substantial scatter, nebular emission is in every case predicted to produce $V-K$ colours much too blue to explain the red excess of BCG halos. 

\begin{figure*}[t]
\epsscale{1.0}
\plottwo{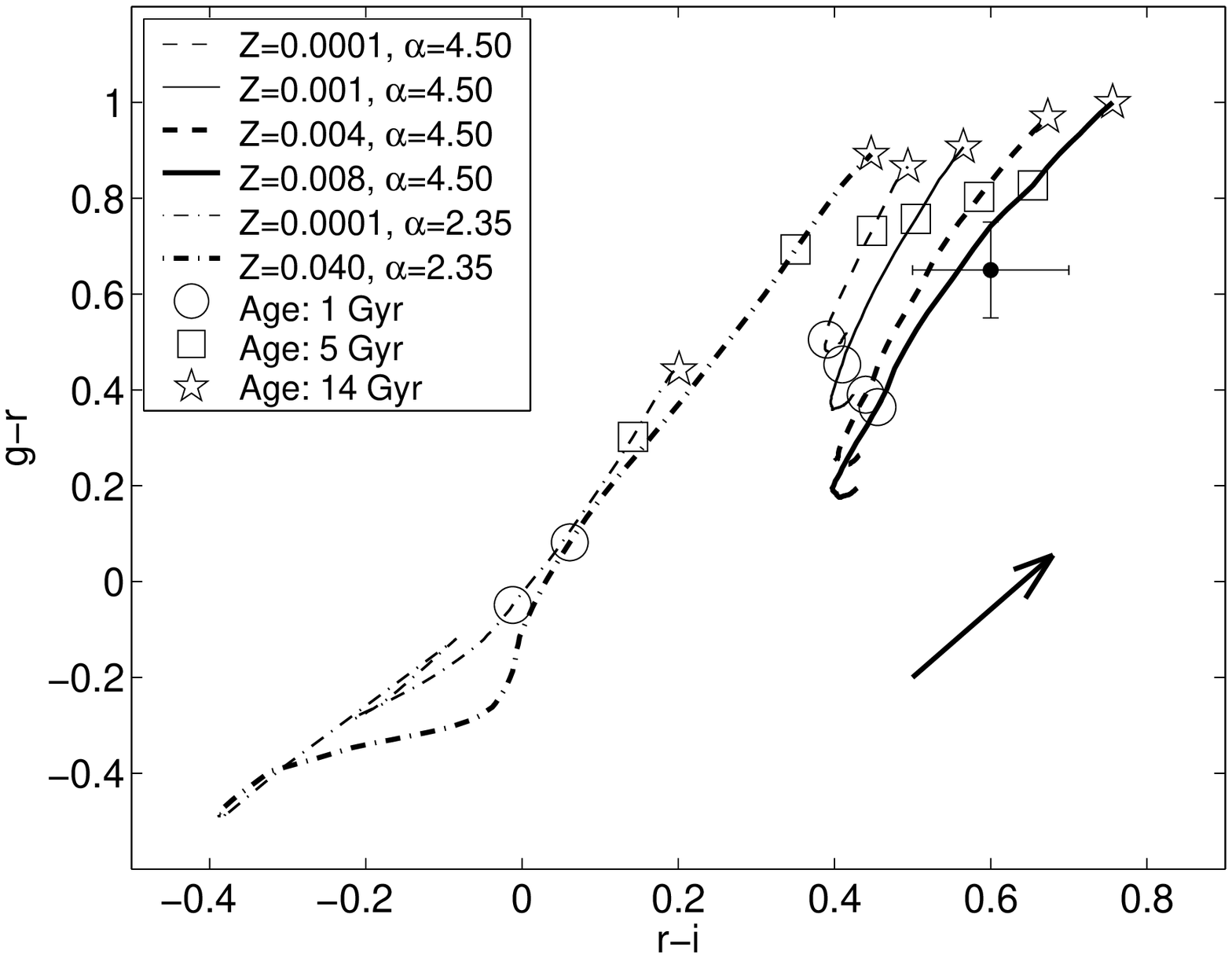}{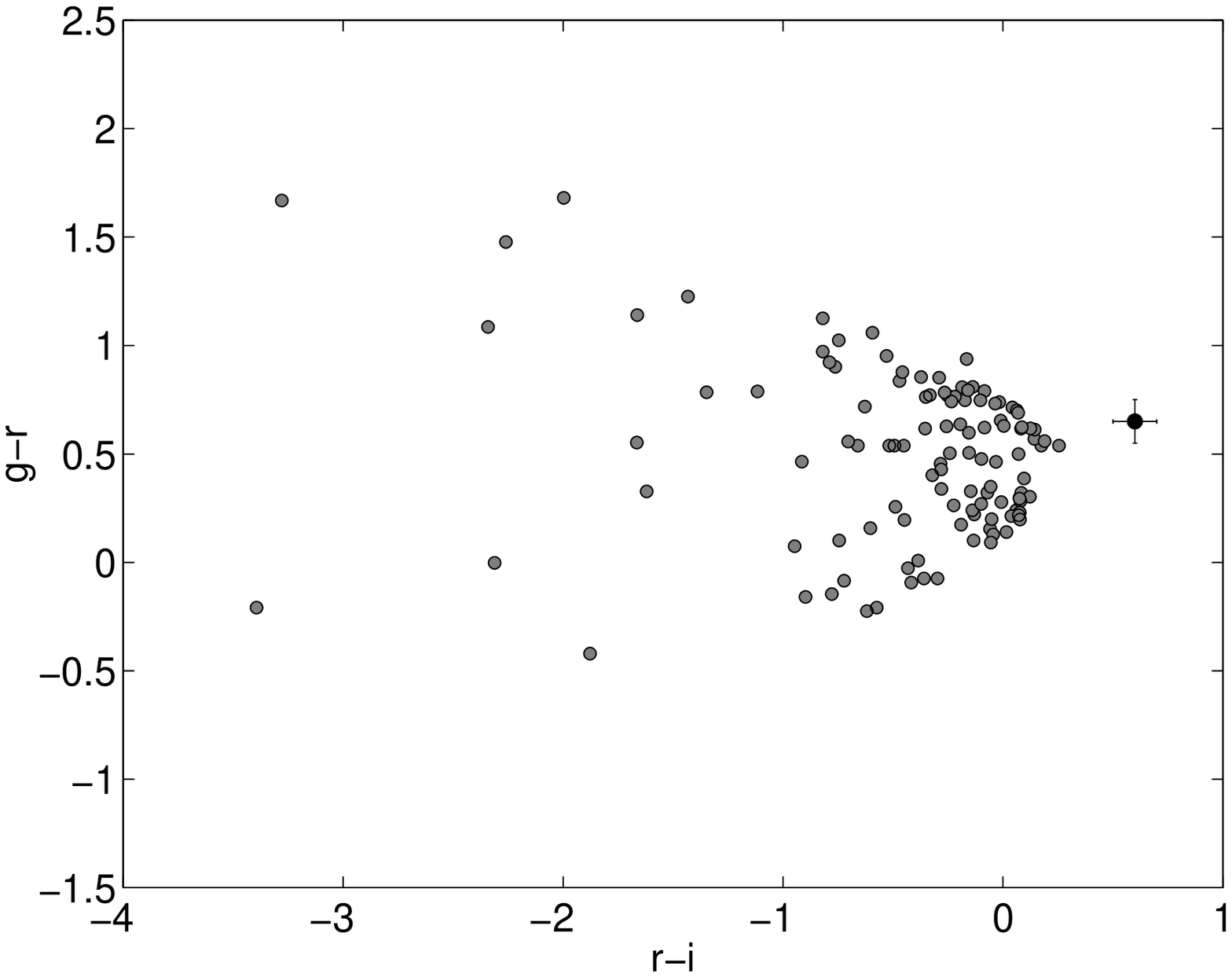}
\caption{{\bf Left:} The observed colour of the halo detected in stacked SDSS edge-on disk galaxy data (cross indicating $1\sigma$ error bar) compared to the predictions of P\'EGASE.2 (lines) for stellar populations with various metallicities and IMF slopes, $\alpha$ ($dN/dM\propto M^{-\alpha}$), as listed in the legend.  Markers along the evolutionary sequences indicate population ages of 1 Gyr (circle), 5 Gyr (square) and 14 Gyr (pentagram). The same star formation history as in Fig.~\ref{BCGfig} has been assumed. The arrow indicates the average Milky Way dust reddening vector \citep{Cardelli et al.} for $E(B-V)=0.25$. The colours of the SDSS halo is well reproduced by an intermediate-metallicity population ($Z=0.004$--0.008; thick dashed, thick solid) with a bottom-heavy IMF ($\alpha=4.50$, $M=0.08$--120 $M_\odot$). By contrast, populations with Salpeter IMFs (dash-dotted lines) fail to reproduce the observed colours, regardless of metallicity. {\bf Right:} Grey markers indicate the colours predicted for various lines of sight through photoionized nebulae. These predictions indicate that nebular emission is much too blue in $r-i$ to explain the observed halo colours. \label{SDSSfig}}
\end{figure*}

For the disk galaxy halos, we use Z01 model spectra for metal-rich ($Z=0.020$) stellar populations which have undergone an exponentially decaying star formation rate (SFR$(t)\propto \exp(-t/\tau)$ with $\tau=6$ Gyr) for 12 Gyr, in agreement with expectations for typical disk galaxies \citep[e.g.][]{Nagamine}. In this case, both high ($Z=0.020$) and low ($Z=0.001$) halo gas metallicities are considered. The resulting line-of-sight colours are confronted with the SDSS halo colours in Fig.~\ref{SDSSfig} (right panel). Since our model predicts nebular emission to be much too blue in $r-i$, ionized gas cannot explain the observed red excess. 

Although we cannot rule out contributions from nebular emission to the observed halos of either BCG or disk galaxies, this component has a spectral signature much too blue to explain the red colours. Correcting the observed halo colours for potential contamination by this blue nebular emission would therefore only make the red excess even more severe.

\begin{figure}
\epsscale{.80}
\plotone{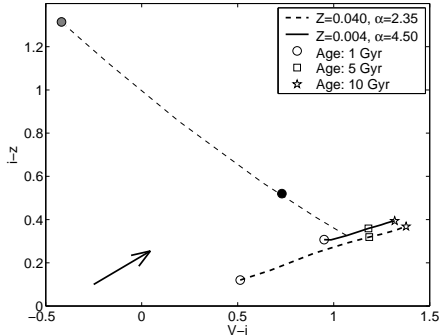}
\caption{The observed colours of the halo detected at $z=0.322$ in the HUDF by \citet{Zibetti & Ferguson} (black marker) compared to the predictions of the Z01 model for an intermediate-metallicity ($Z=0.004$) stellar population with a bottom heavy IMF ($dN/dM\propto M^{-\alpha}$ with $\alpha=4.50$) (thick solid line) and a high-metallicity ($Z=0.040$) Salpeter IMF stellar population (thick dashed line). Neither of these give an adequate fit to the observed colours. Both populations have the same star formation history as those in Fig.~\ref{BCGfig}. Markers along the evolutionary sequences indicate population ages of 1 Gyr (circle), 5 Gyr (square) and 10 Gyr (pentagram; roughly the age of the Universe at this redshift, under the assumption of a $\Omega_\mathrm{M}=0.3$, $\Omega_\Lambda=0.7$, $H_0=72$ km s$^{-1}$ Mpc$^{-1}$ cosmology). The evolution predicted for ages $<1$ Gyr has been excluded in order not to clutter the plot. The grey marker indicates the colour predicted for the spectrum of a nebula with gas metallicity $Z=0.020$, hydrogen number density 100 cm$^{-3}$ and filling factor 0.01, ionized by a $Z=0.020$, $\tau=6$ Gyr ($\mathrm{SFR}(t)\propto \exp{-t/\tau}$) stellar population of age 9 Gyr. As indicated by the dashed line, the observed halo colours can be reproduced by nebular emission superimposed on a $\approx 3$ Gyr old halo population  obeying a bottom-heavy IMF. A similar fit can also be produced by superimposing the nebular emission on the Salpeter-IMF, high-metallicity population. It should be emphasized that these results do not imply that the halo is younger than the disk, as the inferred halo age depends on the assumed star formation history and the adopted gas parameters. The arrow indicates the average Milky Way dust reddening vector \citep{Cardelli et al.} for $E(B-V)=0.25$.
\label{HUDF}}
\end{figure}

The situation is somewhat different for the HUDF halo. As demonstrated in Fig.~\ref{HUDF}, neither a Salpeter-IMF, metal-rich population (thick dashed line) nor an intermediate-metallicity population with a very bottom-heavy IMF (thick solid line) can adequately explain the observed colours. In this case, a component of nebular emission superimposed on either of these stellar populations will however do the trick. Hence, for this particular case, we conclude that contamination by nebular emission is indeed the most likely explanation for the halo colours observed.   

\section{Discussion}
We find that the colours of the red halos detected around BCGs and disk galaxies can be explained by a stellar population of low to intermediate metallicity ($Z=0.001$--0.008) and an extremely bottom-heavy IMF ($\alpha=4.50$).

The IMF slope inferred here is extreme, but similar to that reported for the field population of the LMC \citep[$\alpha\approx 5$;][]{Massey,Gouliermis et al.}. In currently favoured scenarios for galaxy formation, halo stars did not form {\it in situ}, but were shed from small protogalaxies during hierarchical build-up of larger systems \citep[e.g.][]{Abadi et al.}. In this framework, it is conceivable that a halo of low-mass stars could have been created through dynamical mass segregation inside each protogalaxy, followed by tidal stripping of the outer parts of these subunits during the merging process. In alternative scenarios, low-mass stars could for instance have been produced by bottom-heavy star formation in cooling flows \citep[e.g.][]{Mathews & Brighenti}. 

Although the assumption of an IMF overly abundant in low-mass stars allows the observed halo colours to be explained with a lower metallicity than that required by the Salpeter IMF, the inferred metallicity ($Z=0.001$--0.008) is for many of the investigated objects still substantially higher than that of the Milky Way halo \citep[$Z\approx 0.0002$, e.g.][]{Ryan & Norris}. At this time, it is however not completely clear how typical the Milky Way halo really is. Simulations based on hierarchical formation of stellar halos indicate that the halo metallicity is sensitive to the merger history, and that metallicities substantially higher than that of the Milky Way halo can arise in the case of a more protracted assembly \citep{Renda et al.}. Recent observations of red giant branch stars in the halos of nearby disk galaxies indeed suggest that the Milky Way has a metallicity nearly a factor of $\sim~10$ below the average for a galaxy of its luminosity \citep{Mouhcine et al.}

Another curious property of our model fits are the relatively low halo ages  (1--5 Gyr) that are predicted when a SFH typical of an early-type system is assumed (SFR$(t)\propto \exp(-t/\tau)$ with $\tau=1$ Gyr). In principle, the best-fitting age of any single halo can be substantially increased by adopting a more prolonged SFH, but the SFH required to achieve this may not be very realistic. Indeed, to match many of the red halo colours with an $\alpha=4.50$ IMF and a halo age higher than 10 Gyrs seems to require a star formation rate which increases over cosmological time scales. The average SDSS halo colours can for instance be fitted with an age of 13 Gyrs if an exponentially increasing star formation rate with $\tau=-5$ Gyr is adopted. To the best of our knowledge, such an unusual SFH has not been proposed by any halo formation model so far. A more viable explanation is perhaps that current red halo data still suffer from contamination by light from young stars, nebular emission, or some combination thereof. While such contamination cannot explain the red excess itself (as discussed in in Sect. 5), it may nonetheless have shifted the halo colours in the blueward direction. Because of its more realistic treatment of nebular emission, we turn to the Z01 model to explore these possibilities in detail. 

In the case of BCGs, there is no doubt ample supply of young stars in the central starburst, and diffuse nebular emission is in many cases detectable out to considerable distances from the centre \citep[e.g.][]{Izotov et al.,Guseva et al.}. Assuming that the uncontaminated red halo is well-described by a $Z=0.004$, 13 Gyr old population with an $\alpha=4.50$, 0.08--120 $M_\odot$ IMF, we can estimate the contamination required from a young (100 Myr), low-metallicity ($Z=0.001$) Salpeter-IMF population with associated nebular emission in order to shift the best-fitting halo age below 5 Gyr. We find, that to achieve this, the young populations would have to contribute 60--80\% of the light to the $B$-band flux in the region where the BCG red halo colours are measured. Because of the radically different $M/L$-ratios of these two populations, the starburst component would however only need to make up 0.01--0.1\% of the stellar mass in this region. Since the shift essentially follows the predicted age vector of the $\alpha=4.50$ population, correcting for this contamination would increase the red halo age while leaving the best-fitting metallicity and IMF slope intact. Quantitative predictions for the scenario in which the $\alpha=4.50$ halo is mixed with pure nebular emission are not as straightforward to derive, due to the large scatter in colours among the synthetic nebulae (Fig.~\ref{BCGfig}, right panel). It is clear, however, that to affect the halo $B-V$ colours, most of the nebular component models would produce an even larger shift in $V-K$. By adopting a nebular component model with average colours from the ensemble plotted in Fig.~\ref{BCGfig}, we find that around 80\% of the measured $B$-band flux in the halo region would have to stem from the ionized gas in order to make the measured halo $B-V$ consistent with an age of 13 Gyr for the $\alpha=4.50$ halo. Correcting for this contamination would however (because of the simultaneous shift in $V-K$ by at least 1 magnitude) push the best-fitting halo model to $Z>0.008$ and/or $\alpha>4.50$. 

While normal disk galaxies do not typically feature prominent starbursts, they do display a certain degree of active star formation. By considering flux contamination from a 12 Gyr old, $Z=0.020$, $\tau=6$ Gyr (SFR$(t)\propto \exp{-t/\tau}$) disk population with associated nebular emission, the Z01 code predicts that a $Z=0.004$, $\alpha=4.50$, 13 Gyr old halo indeed could appear to have an age below 5 Gyr, provided that the flux contribution from the disk amounts to at least 60\% in $g$-band. In terms of stellar mass, this corresponds to a 2\% contribution from the disk population in the region where the halo colours are measured. Analogous to the case for BCGs, this additional component mainly shifts the colours along the age vector of the underlying halo component, without substantially affecting the best-fitting halo metallicity or IMF slope. Contamination by the pure nebular emission associated with this disk model would, however, mainly change the $r-i$ colour, while leaving the more age-sensitive $g-r$ colour largely unaffected. Hence, such a mixture could not possibly mimic a young halo age. 

Since contamination by light associated with the central starburst or disk population could in principle have affected the red halo colours measured, constraining the contribution from such additional components to the spectal energy distribution of the halo therefore constitutes an important challenge for future red halo observations. A possible way to test for the presence of ionizing components such as these would be to carry out narrowband imaging of the halo in filters centred on strong emission lines.  For BCGs, we predict that the equivalent width of H$\alpha$ in emission would have to be $\mathrm{EW}(\mathrm{H}\alpha)=300$--500 \AA{} in the halo if stars and nebular emission have indeed affected the inferred ages in the way considered here. If the contamination is by nebular emission alone, the corresponding value would be $\mathrm{EW}(\mathrm{H}\alpha)> 1000$ \AA. Because of the more modest star formation activitiy in disk galaxies, the mixture used in our calculations would on the other hand only result in $\mathrm{EW}(\mathrm{H}\alpha)\approx 30$ \AA{} for the red halos of such objects.

If halos around galaxies of very different Hubble types indeed contain substantial stellar populations dominated by low-mass stars, these stars may contribute to the baryonic dark matter content of the Universe, of which $\sim 1/3$ is still missing \citep{Fukugita}. While it has been argued that the missing baryons should mainly be in the form of a warm-hot intergalactic medium \citep[e.g.][]{Dave et al.} or in hot gas around galaxies \citep[e.g.][]{Fukugita & Peebles,Sommer-Larsen}, the errorbars on current observations of such components \citep[e.g.][]{Nicastro et al.,Pedersen et al.} are still too large to robustly rule out additional reservoirs of dark baryons. The $M/L$-ratio of the stellar population postulated here as a solution to the red halo problem mainly depends on the assumed lower mass limit of the IMF. As discussed in Sect. 3, the red halo colours can be reproduced by the models if an $\alpha=4.50$ power-law extending at least down to $0.1 \ M_\odot$ is assumed. This implies a halo population mass-to-light ratio of $M/L_B \gtrsim 40$, which in all but the very deepest images would qualify as dark matter. A stellar halo of this type would contribute substantially to the baryon budget of galaxies, even if it contributes only a few percent to the total luminosity. 

The possibility of a stellar halo with an extremely bottom-heavy IMF as an explanation for the red halo phenomenon in disk galaxies and BCGs can be subjected to a number of additional tests. 

In the case of disk galaxies, a combination of rotation curve data and halo photometry can be used to test whether a population with $M/L_B \gtrsim 40$ is at all plausible, since the mass density provided by the red halo stellar population must at all distances from the centre be lower (and probably substantially so) than the overall mass density. 

For the BCGs, a metal-rich halo stellar population can be distinguished from the suggested $\alpha=4.50$ population by the addition of deep $I$-band data, which is currently lacking, since these two populations occupy completely disconnected parts of a diagram of e.g. $V-I$ vs. $B-V$. Due to the low surface brightness of the regions from which the current $B-V$ and $V-K$ halo colours are derived, the error bars on these quantities are also quite large. To improve the situation, we are currently developing more sophisticated image processing techniques (Micheva et al., in preparation), which will allow us to decrease the photometric uncertainties and impose tighter constraints on the different scenarios (\"Ostlin et al., in preparation).

A halo population dominated by low-mass stars should furthermore be detectable by Spitzer in the mid-IR. By correlating rotation curves with Spitzer data, it should therefore be possible to constrain the contribution of the red halo to the total mass budget of the galaxy, using an approach similar to that of \citet{Gilmore & Unavane} for ISO data.

At some point, it may also be interesting to address the issue of whether a substantial population of low-mass stars could be present in the halo of the Milky Way as well. The MACHO and EROS/EROS2 projects \cite[e.g.][]{Alcock a, Lasserre et al.,Tisserand & Milsztajn} have already ruled out such objects as the main constituent of the halo dark matter, as would be expected in a universe whose matter sector is dominated by non-baryonic dark matter. For a population of hydrogen-burning stars of the type proposed here, even stronger constrains may however be imposed by star counts in deep fields  \citep[see e.g.][ and references therein]{Brandner}.

\section{Summary}
We suggest a stellar population of low to intermediate metallicity and an extremely bottom-heavy IMF as an explanation for the red integrated halo light detected in very deep observations of both BCGs and edge-on disk galaxies. Other previously suggested explanations, like nebular emission or very metal-rich stars, can on the other hand only explain the red halo excess in certain objects. This halo population has a sufficiently high mass-to-light ratio to qualify as baryonic dark matter.


\begin{thebibliography}{}
\bibitem[Abadi et al.(2006)Abadi, Navarro, \& Steinmetz]{Abadi et al.}
Abadi, M. G., Navarro, J. F., \& Steinmetz, M. 2006, MNRAS, 365, 747
\bibitem[Alcock et al.(1998)]{Alcock a}
Alcock, C., et al. 1998, ApJ, 499, L9
\bibitem[Alcock et al.(2000)]{Alcock b}
Alcock, C., et al. 2000, ApJ, 542, 281
\bibitem[Bergvall \& \"Ostlin(2002)]{Bergvall & Östlin}
Bergvall, N., \& \"Ostlin, G. 2002, A\&A, 390, 891
\bibitem[Bergvall et al.(2003)]{Bergvall et al.}
Bergvall, N., Marquart, T., Persson, C., Zackrisson, E., \& \"Ostlin, G. 2003, in Bender, R. Renzini, A., eds, Multiwavelength Mapping of Galaxy Formation and Evolution, Springer-Verlag, in press
\bibitem[Brandner(2004)]{Brandner}
Brandner, W. 2004, astro-ph/0411276
\bibitem[Bruzual \& Charlot(2003)]{Bruzual & Charlot}
Bruzual, G., \& Charlot, S. 2003, MNRAS, 344, 1000
\bibitem[Calchi Novati et al.(2005)]{Calchi Novati}
Calchi Novati, S., et al. 2005, astro-ph/0504188
\bibitem[Cardelli et al.(1989)]{Cardelli et al.}
Cardelli, J. A., Clayton, G. C.,\& Mathis, J. S. 1989, ApJ, 345, 245
\bibitem[Dalcanton \& Bernstein(2002)]{Dalcanton & Bernstein}
Dalcanton, J. J., \& Bernstein, R. A. 2002, AJ, 124, 1328
\bibitem[Dav\'e et al.(2001)]{Dave et al.}
Dav\'e, R., et al. 2001, ApJ, 552, 473
\bibitem[Ferland et al.(1998)]{Ferland et al.}
Ferland, G. J., Korista, K. T., Verner, D. A., Ferguson, J. W., Kingdon, J. B., \& Verner, E. M. 1998, PASP, 110, 761
\bibitem[Fioc \& Rocca-Volmerange(1999)]{Fioc & Rocca-Volmerange}
Fioc, M., \& Rocca-Volmerange, B. 1999, astro-ph/9912179
\bibitem[Fukugita(2004)]{Fukugita}
Fukugita, M. 2004, in  S. D. Ryder, D. J. Pisano, M. A. Walker, and K. C. Freeman, eds, International Astronomical Union Symposium no. 220, Astronomical Society of the Pacific., p.227
\bibitem[Fukugita \& Peebles(2006)]{Fukugita & Peebles}
Fukugita, M., \& Peebles, P. J. E. 2006, ApJ, 639, 590
\bibitem[Gilmore \& Unavane(1998)]{Gilmore & Unavane}
Gilmore, G., \& Unavane, M. 1998, MNRAS, 301, 813
\bibitem[Gouliermis et al.(2006)]{Gouliermis et al.}
Gouliermis, D., Brandner, W., Henning, T., 2006, ApJ, 641, 838
\bibitem[Guseva et al.(2004)]{Guseva et al.}
Guseva, N. G., Papaderos, P., Izotov, Y. I., Noeske, K. G., \& Fricke, K. J. 2004, A\&A, 421, 519\\
\bibitem[Izotov et al.(2001)]{Izotov et al.}
Izotov, Y. I., et al. 2001, A\&A 378, 756
\bibitem[James \& Casali(1998)]{James & Casali}
James, P. A., \& Casali, M. M. 1998, MNRAS, 301, 280
\bibitem[Lasserre et al.(2000)]{Lasserre et al.}
Lasserre, T., et al. 2000, A\&A, 355, 39
\bibitem[Lequeux et al.(1996)]{Lequeux et al. a}
Lequeux, J., Fort, B., Dantel-Fort, M., Cuillandre, J.-C., \& Mellier, Y. 1996, A\&A, 312, L1
\bibitem[Lequeux et al.(1998)]{Lequeux et al. b}
Lequeux, J., Combes, F.,  Dantel-Fort, M., Cuillandre, J.-C., Fort, B.,\& Mellier, Y. 1998, A\&A, 334, L9
\bibitem[Massey(2002)]{Massey}
Massey, P. 2002, ApJS, 141, 81
\bibitem[Mathews \& Brighenti(1999)]{Mathews & Brighenti}
Mathews, W. G., \& Brighenti, F. 1999, ApJ, 526, 114
\bibitem[Michard(2002)]{Michard}
Michard, R. 2002, A\&A, 384, 763
\bibitem[Mouhcine et al.(2005)]{Mouhcine et al.}
Mouhcine, M., Ferguson, H. C., Rich, R. M., Brown, T. M., \& Smith, T. E. 2005, ApJ, 633, 821
\bibitem[Nagamine et al.(2001)]{Nagamine}
Nagamine, K., Fukugita, M., Cen., R., \& Ostriker, P. O. 2001, ApJ, 558, 497
\bibitem[Nicastro et al. (2005)]{Nicastro et al.}
Nicastro, F., et al. 2005, Nature, 433, 495
\bibitem[\"Ostlin et al.(2001)]{Östlin et al.}
\"Ostlin, G., Amram, P., Bergvall, N., Masegosa, J., Boulesteix, J., \& M\'arquez, I. 2001, A\&A, 374, 800
\bibitem[Pedersen et al.(2006)]{Pedersen et al.}
Pedersen, K., Rasmussen, J., Sommer-Larsen, J., Toft, S., Benson, A. J., Bower, R. G. 2006, New Astronomy 11, 465 
\bibitem[Renda et al.(2005)]{Renda et al.}
Renda, A., Gibson, B. K., Mouhcine, M., Ibata, A. R., Kawata, D., Flynn, C., \& Brook, C. B. 2005, astro-ph/0507281
\bibitem[Rudy et al.(1997)]{Rudy et al.}
Rudy, R. J., Woodward, C. E., Hodge, T., Fairfield, S. W., \& Harker, D. 1997, Nature, 387, 159
\bibitem[Ryan \& Norris(1991)]{Ryan & Norris}
Ryan, S. G., \& Norris, J. E. 1991, AJ, 101, 1865
\bibitem[Sackett et al.(1994)]{Sackett et al.}
Sackett, P. D., Morrison, H. L. Harding, P., \& Boroson, T. A. 1994, Nature, 370, 441
\bibitem[Shang et al.(1998)]{Shang et al.}
Shang, Z., et al. 1998, ApJ, 504, 23
\bibitem[Sommer-Larsen(2006)]{Sommer-Larsen}
Sommer-Larsen, J. 2006, astro-ph/0602595
\bibitem[Tisserand \& Milsztajn(2005)]{Tisserand & Milsztajn}
Tisserand, P., \& Milsztajn, A. 2005, astro-ph/0501584
\bibitem[Zackrisson et al.(2001)]{Zackrisson et al.}
Zackrisson, E., Bergvall, N., Olofsson, K., \& Siebert, A., 2001, A\&A, 375, 814 (Z01)
\bibitem[Zepf et al.(2000)]{Zepf et al.}
Zepf, S. E., Liu, M. C., Marleau, F. R., Sackett, P. D., \& Graham, J. R. 2000, AJ, 119, 1701
\bibitem[Zheng et al.(1999)]{Zheng et al.}
Zheng, Z., et al. 1999, AJ, 117, 2757
\bibitem[Zibetti et al.(2004)Zibetti, White, \& Brinkmann]{Zibetti et al.}
Zibetti, S., White, S. D. M., \& Brinkmann, J. 2004, MNRAS, 347, 556
\bibitem[Zibetti \& Ferguson(2004)]{Zibetti & Ferguson}
Zibetti, S., \& Ferguson, A. M. N. 2004, MNRAS, 352, 6
\end{thebibliography}
\end{document}